\documentclass{tPHL2e}

\usepackage{epstopdf}
\usepackage{subfigure}

\theoremstyle{plain}

\theoremstyle{definition}

\theoremstyle{remark}

\begin{document}



\title{\textit{On the electronic viscosity of a Dirac fluid in deformed graphene}}

\author{
\name{Sergei Sergeenkov\textsuperscript{a}$^{\ast}$\thanks{$^\ast$Corresponding author. Email: sergei@df.ufscar.br} and  Marcel Ausloos\textsuperscript{b}}
\affil{\textsuperscript{a}Departamento de F\'{\i}sica, Universidade Federal da Para\'{\i}ba, Jo\~{a}o Pessoa, PB, Brazil; \\
\textsuperscript{b}Group for Research on Applications of Physics in Economy and Sociology (GRAPES), rue de la Belle Jardiniere, 483/021, B-4031 Liege Angleur, Belgium}}

\maketitle

\begin{abstract}
We discuss the properties of the electronic viscosity of a Dirac fluid in deformed graphene by introducing  a strain $\epsilon$ and a velocity gradient $\nabla v$, as  equivalent to a pseudo-magnetic $B^d \propto \hbar \epsilon$ and a pseudo-electric $E^d \propto \hbar \nabla v$ field respectively into the Dirac equation. It is thereby analytically established that the dynamic shear viscosity coefficient $\eta$ substantially decreases with the applied strain as $\eta (\epsilon)=\eta (0)(1+2\epsilon)^{-3/2}$, reaching as much as $[\eta (0)-\eta (\epsilon)]/\eta (0)\simeq 70\%$ for $\epsilon=0.5$.
\end{abstract}


\section{Introduction}

Still full of surprises, graphene   continues to attract attention as a unique test ground for many interesting ideas which are expected to convert into new technological applications of this unique material (see,  e.g.~\cite{i1,i2,i3,i4,i5} and many references therein). In particular, graphene electronic properties have been shown to closely follow predictions of the relativistic hydrodynamics with extremely small value of the so called Dirac fluid viscosity~\cite{a1,a2,a3}. This phenomenon, coined "an almost perfect electronic fluid", has been attributed to manifestation of specific topological properties of graphene. Moreover, many interesting and unusual phenomena in graphene under mechanical deformations (leading to substantial modifications of its electronic structure) have been recently observed or predicted ~\cite{1a,2a,3a,4a,5a,5b,5c,6a,7a,7b,8a}. In particular, it has been established ~\cite{1a,2a}  that the homogeneous strain $\epsilon$ induced gauge potential $A^d_y=B^d_zx$ results in a pseudo-magnetic field $B^d_z=\hbar {\epsilon} /er^2$ inside deformed graphene (with $r=0.14nm$ being the carbon-carbon bond length). 

In this Letter, we discuss the properties of  the dynamic viscosity $\eta$ of the electronic fluid   in a strained graphene,  by introducing a homogeneous strain $\epsilon$ and a constant velocity gradient $\nabla v$ into the Dirac  equation via a pseudomagnetic and a pseudoelectric field, respectively. It is shown that the obtained analytical results suggest quite a realistic possibility for further "improvement" of the "perfectness" of the Dirac fluid in a strained graphene.

\section{Results and Discussion}

Recall that in the absence of chirality (intervalley) mixing ~\cite{1a,2a,5c}, the low-energy electronic properties of graphene near the Fermi surface can be reasonably described by a two-component wave function $|\Psi>=(\Psi_A, \Psi_B)$ obeying a
massless Dirac equation
\begin{equation} \label{eq1}
{\cal H}|\Psi>=E|\Psi>
\end{equation}\
with an effective Hamiltonian
\begin{equation} \label{eq2}
{\cal H}=v_F(\sigma_x \pi_x+\sigma_y \pi_y)+eV_d
\end{equation}
Here, $\pi_a=p_a+eA^d_a$ with $p_a=-i\hbar \nabla_a$ being the
momentum operator and $A^d_a=(0,A^d_y)$ the deformation induced
vector potential, $eV_d$ is a chemical potential; $\sigma_a$ are
the Pauli matrices, and $v_F$ is the Fermi velocity. In what
follows, $a =\{x,y\}$.

By analogy with plastically deformed graphene ~\cite{5c},  viscosity prone effects are introduced  into the model through a constant velocity gradient $\nabla_y v$ induced scalar potential $V_d=E^d_yy$ where $E^d_y=\hbar \nabla_y v/er$ is a
pseudo-electric field created by a moving electronic fluid.
Without losing generality, in what follows we assume that the
velocity $v$ is a scalar and use $\nabla_y v\equiv \nabla v$ to
simplify notations. 
It can be easily verified that Dirac equation Eq.(\ref{eq1}) with the
above-defined potentials have the normalized solutions
\begin{equation}\label{eqPsiA}
\Psi_A=C_1\exp\left[i(k_1x+k_2y)-\frac{(x-x_0)^2}{2l_1^2}-\frac{(y+y_0)^2}{2l_2^2}\right]
\end{equation}
and
\begin{equation} \label{eqPsiB}
\Psi_B=C_2\left[i\frac{(x-x_0)}{l_1^2}+\frac{(y+y_0)}{l_2^2}\right]\Psi_A
\end{equation}
with $k_1=x_0/l_1^2$, $k_2=y_0/l_2^2$, and the total energy
\begin{equation}\label{eq5}
E=\hbar v_F\sqrt{\frac{2}{l_1^{2}}+\frac{2}{l_2^{2}}+\frac{2}{A}}
\end{equation}
where $l_1^2=\hbar/eB^d_z$, $l_2^2=\hbar v_F/eE^d_y$, and $A=L^2$ is a square sheet area of graphene.

Recall that the linear, i.e. $\nabla v$ independent, coefficient
of the dynamic shear viscosity $\eta$ is defined via  the stress field
$\sigma$ created by a gradient of velocity $v$ as
\begin{equation} \label{eq6}
\sigma=\eta \nabla v
\end{equation}
On the other hand, the stress field can be generally defined as a
strain $\epsilon$ induced response of the electronic fluid, namely
\begin{equation} \label{eq7}
\sigma \equiv -\frac{1}{A}\left [\frac{\partial E}{\partial
\epsilon}\right ]
\end{equation}

\begin{figure}
\centerline{\includegraphics[width=10cm]{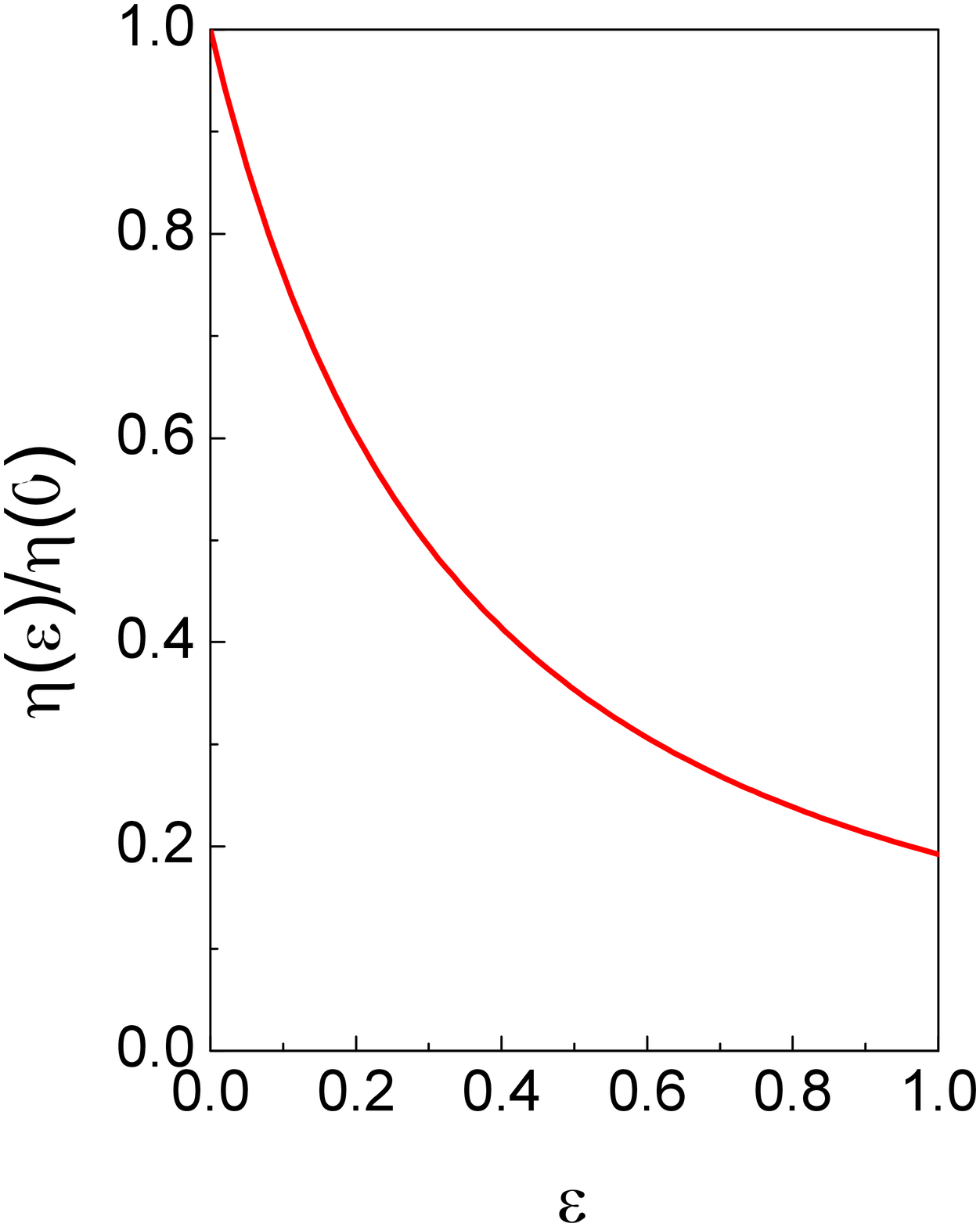}}\vspace{0.25cm}
\caption{The strain-induced behavior of viscosity coefficient according to Eq.(8). }
\label{fig:fig1}
\end{figure}

As a result, from Eqs.(\ref{eq5})-(\ref{eq7}), we obtain 
\begin{equation} \label{eq8}
\eta (\epsilon)\equiv \left [\frac{\partial \sigma}{\partial \nabla v}\right
]_{\nabla v=0}=\eta (0) \left(\frac{1}{1+2\epsilon} \right)^{3/2}
\end{equation} 
for the coefficient of shear viscosity in strained graphene where $\eta (0)=\hbar/L^{2}$ is the strain-free value.  
Fig.\ref{fig:fig1} shows the decrease of the viscosity coefficient with the increase of strain  $\epsilon$. In order to compare the hereby  predicted   effects with the available
data based on a hydrodynamic approach \cite{a1,a2,a3}, let us introduce the so-called kinematic viscosity $\nu$ which is related to the shear coefficient via the mass density $\rho$ of the fluid as follows, $\nu =\eta /\rho$. According to \cite{a3}, at room
temperature the mass density of electronic fluid in graphene is of the order of $\rho \simeq 6\times 10^{-19}kg/m^2$ leading to $\nu(0) \simeq 0.005m^2/s$.
It can be easily verified that this value of $\nu(0)$ corresponds to $L=0.1\mu m$. At the same time, according to  the above,  an application of an experimentally achievable strain \cite{b} of $\epsilon=0.5$ will result in a significant drop of the kinematic viscosity, reaching $[\eta (0)-\eta (\epsilon)]/\eta (0)\simeq 70\%$ (see Fig.1). It is worthwhile to mention that in graphene electron fluid velocities can reach as high as \cite{c} $v\simeq 0.1v_F$ leading to gradients $\nabla v\simeq v/L\simeq 0.1v_F/L\simeq 10^{12}s^{-1}$ for $L=0.1\mu m$.

\section{Conclusions}

In conclusion,  let us emphasize that the conventional electron-electron interaction usually considered for calculating  the shear viscosity coefficient \cite{d}  is replaced here by introducing a velocity gradient equivalent to a pseudo-gauge field  directly  into the Dirac equation. It is demonstrated that the electronic viscosity of the Dirac fluid in graphene can be substantially reduced under strong enough mechanical deformation. The obtained results are expected to be relevant for potential nanoelectronics applications \cite{a1,e}.

\section*{Acknowledgements}

We are indebted to Igor Barashenkov (Cape Town) and Yury Shukrinov (Dubna) for very useful discussions. This work has been financially supported by the Brazilian agency FAPESQ (DCR-PB).

\end{document}